%% file: main.tex
\documentclass[12pt]{article}
\usepackage[utf8]{inputenc}
\usepackage[left=3cm, right=3cm, top=2cm]{geometry}
\usepackage{graphicx}
\usepackage{siunitx}
\usepackage{booktabs}
\usepackage{rotating}
\usepackage{multirow}
\usepackage{authblk}

\newcommand{\drm}{\mathrm{d}}

\title{Measurement of  deuteron carbon vector analyzing powers in the kinetic energy range 170 - 380 \si{MeV}}

\begin{document}

\input{authors}
\maketitle
\begin{center}
    (JEDI collaboration)
\end{center}

\begin{abstract}
A measurement of vector analyzing powers in elastic deuteron-carbon scattering has been performed at the Cooler Synchrotron COSY of Forschungszentrum J\"ulich, Germany.
Seven kinetic beam energies between 170 and 380 MeV have been used.
A vector-polarized beam from a polarized deuteron source was injected,
accelerated to the final desired energy and stored in COSY. A thin needle-shaped diamond strip was used as a carbon target, onto which the beam was slowly steered. Elastically scattered deuterons were identified in the forward direction using various layers of scintillators and straw tubes. 

Where data exist in the literature (at 200 and 270 MeV), excellent agreement of the angular shape was found. The beam polarization of the presented data was deduced by fitting the absolute scale of the analyzing power to these references. Our results extend the world data set and are necessary for polarimetry of future electric dipole moment searches at storage rings. They will as well serve as an input for theoretical description of polarized hadron-hadron scattering.

\end{abstract}

\section{Introduction}
The main motivation for the measurements arise from plans to search for 
electric dipole moments (EDMs) of charged hadrons in storage rings~\cite{Abusaif:2019gry}.
The existence of electric dipole moments is connected to CP-violation and is therefore closely related to fundamental questions of the dominance of matter over anti-matter in the universe~\cite{Bernreuther:2002uj}.  A storage ring EDM  measurement is based on the observation of a vertical polarization build-up of an initially horizontal polarization. 
This build-up is proportional to the EDM itself and to the analyzing power 
of the scattering process used to determine the beam polarization. 
The statistical error of the EDM measurement is inversely proportional
the absolute value of the analyzing power.
It is thus important to know the analyzing power in a wide energy range. 

The paper is organized as follows. Section~\ref{sec:mot}
outlines the theoretical background and gives an overview of existing data. Section~\ref{sec:exp} explains the experimental setup.
The analysis and results are presented in section~\ref{sec:analysis}.

\section{Theoretical Background \& Motivation}\label{sec:mot}
\begin{figure}[h!]
    \centering
    \includegraphics[width=0.7\textwidth]{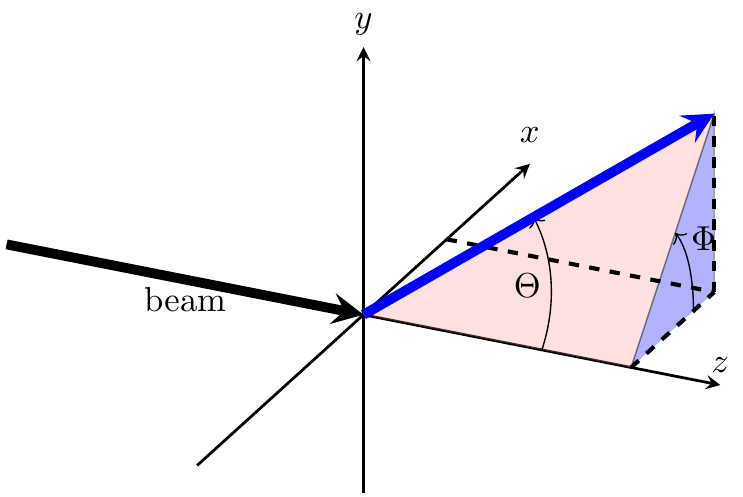}
    \caption{{Definition of the laboratory coordinate system
    with the $z-$axis along the beam direction.
    The blue arrow indicates the direction of the scatterd deuteron.}}
    \label{fig:cs}
\end{figure}

The elastic cross section for purely vector polarized deuterons scattering from a spin 0 target like carbon is given by~\cite{Ohlsen:1973wf}:
\begin{equation}
    \left(\frac{\drm \sigma_{pol.}^{dC}(\Theta, \Phi)}{\drm
    \Omega}\right) = \left(\frac{\drm\sigma_0(\Theta)}{\drm\Omega}\right)\cdot\left(1 + \frac{3}{2}A_y(\Theta)P_y\cos(\Phi)\right),
    \label{eq:pol_cs}
\end{equation}
where the polar and azimuthal angles of the detected particles in the laboratory system are denoted by $\Theta$ and $\Phi$, respectively, see Figure~\ref{fig:cs}.
The vector analyzing power $A_y(\Theta)$ is a property of the elastic scattering $^{12}\mathrm{C(d,d)}^{12}\mathrm{C}$ between the carbon target and the polarized deuteron beam. 
The projection of the polarization vector along the vertical axis is denoted by $P_y$.

Due to azimuthal symmetry, the unpolarized deuteron carbon cross section $\left(\frac{d\sigma_0(\Theta)}{d\Omega}\right)$  depends only on the polar angle $\Theta$.
The number of detected scattered particles is given by:
\begin{equation}\label{eq:N1}
    \frac{\drm N(\Theta, \Phi)}{\drm \Omega} = \mathcal{L}\cdot\alpha(\Theta, \Phi)\cdot\left(\frac{\drm \sigma_0(\Theta)}{\drm\Omega}\right)\cdot\left(1 + \frac{3}{2}A_y(\Theta)P_y\cos(\Phi)\right),
    \label{eq:N_scattered}
\end{equation}
where $\mathcal{L}$ denotes the integrated luminosity and $\alpha(\Theta, \Phi)$
the detector acceptance. Equation~\ref{eq:N1} is the basis for the extraction of the vector analyzing power as described in Section~\ref{sec:analysis}.

From previous measurements at $T=200$~\cite{art:kawabata} and $270\,\si{MeV}$~\cite{art:satou}
it is known that the unpolarized differential cross section of elastic  deuteron carbon 
scattering at a few hundred MeV has a pronounced diffractive structure with some minima and maxima in the forward hemisphere.
    Furthermore, the vector analyzing power of this reaction, $A_y$,  has a maximum value close to unity.
    As a result, this reaction is ideally suited for precision polarization measurements. 
     Therefore, high quality  experimental data on the $A_y$ 
for $\mathrm{d}\,^{12}\mathrm{C}$ elastic scattering
    in a broad range of kinetic energy and scattering angle of the 
deuteron are of importance.

    Available data on unpolarized differential cross sections of  $\mathrm{d}\,^{12}\mathrm{C}$ elastic scattering 
    can be found at $T =$94 MeV, 125 MeV, 156 MeV \cite{Baldw56}, at 110 MeV and 120 MeV 
    \cite{Betker:1993zz},
    at 170 MeV \cite{Baumer:2001ed},  425 MeV \cite{Button1960}, and  650 MeV \cite{Dutton:1965xra}.
    Theoretical  interpretation  of this data at high energy of 650 MeV was done within the Glauber multiple 
    scattering  theory for nucleus-nucleus scattering  using the harmonic oscillator  shell model wave functions
    for the $^{12}\mathrm{C}$ nucleus
    and using the nucleon-nucleon data~\cite{Chadha:1976vh}, or      nucleon-nucleus scattering amplitudes~\cite{Varma:1977pu},
    and deuteron proton scattering data \cite{Ghosh:1978ez} as input. At lower energies  
    $T$=94 MeV, 125 MeV,
    156 MeV the Glauber theory was applied 
    to the $\mathrm{d}\,^{12}\mathrm{C}$ elastic scattering in \cite{ElGogary:2003jd}.

     Experimental data on polarized  d$^{12}$C elastic scattering in the considered
     region of energy are  poorer. In reference~\cite{art:satou}  the vector $A_y$ and tensor $A_{yy}$ analyzing powers
     and the unpolarized differential cross section were  measured  in elastic and inelastic
     $\mathrm{d}\,^{12}\mathrm{C}$ scattering at $T=270$ MeV.  Data at $T=200\,\si{MeV}$ are discussed in reference~\cite{art:kawabata}.

         The aim of this paper is the measurement of the vector analyzing power $A_y$ of  
         $\mathrm{d}\,^{12}\mathrm{C}$ elastic scattering at several kinetic energies in the region $T=170-380$ MeV in the forward
         hemisphere. In addition to the use of this data for polarimetry in storage EDM experiments, it can be used
         for a better understanding of the dynamics of the $\mathrm{d}\,^{12}\mathrm{C}$ elastic scattering and its connection
         to the properties of the $^{12}\mathrm{C}$ nucleus and NN-, dN-, and  $^{12}\mathrm{C}\mathrm{N}$-scattering amplitudes, where N denotes the
         proton or neutron.

\section{Experimental Setup}\label{sec:exp}
\begin{figure}
    \centering
    \includegraphics[width=0.8\textwidth]{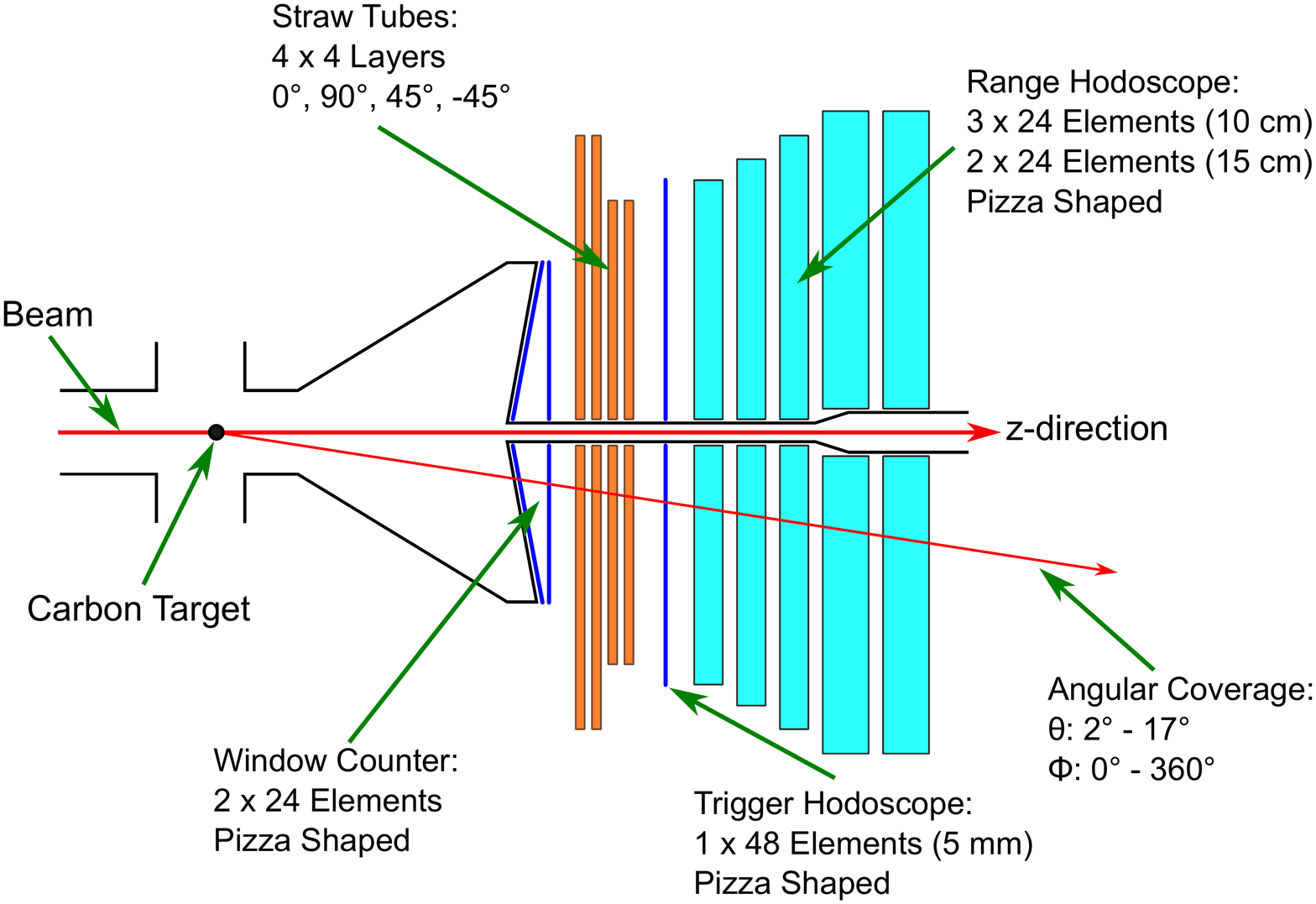}
    \caption{Schematic cross section of the forward part of the WASA detector. It is composed of two layers of thin plastic scintillators called FWC1 and FWC2 (\textit{\textbf{F}orward \textbf{W}indow \textbf{C}ounter}) whose main purpose together with the FTH (\textit{\textbf{F}orward \textbf{T}rigger \textbf{H}odoscope}) is to generate the trigger signal for the data acquisition system. The four successively rotated layers of straw tubes provide the angular information for each track. The five layers of thick plastic scintillators called FRH1 to FRH5 (\textit{\textbf{F}orward \textbf{R}ange \textbf{H}odoscope}) are used to obtain the energy information for each track. }
    \label{fig:WASA}
\end{figure}

The  measurements were performed at the COSY accelerator facility at the \textit{Forschungszentrum J\"ulich} in Germany~\cite{Maier:1997zj}. A pure vector polarized deuteron beam was produced~\cite{Chiladze:2005qw}, accelerated and stored in the COSY ring. Seven different beam energies (180~MeV, 200~MeV, 235~MeV, 270~MeV, 300~MeV, 340~MeV, and 380~MeV) were used. For the extraction of the vector analyzing power, the beam polarization was set to cycle through three polarization states: Unpolarized ($P^0$), upwards polarized {in the laboratory frame} along the vertical axis 
($P_y=P^\uparrow$), and downwards polarized along the vertical axis ($P_y=-P^\downarrow$). The absolute polarization value in the vertical state had to be determined later, see Section~\ref{sec:analysis}. 
For data taking, the beam was slowly steered onto a thin, needle-shaped diamond target to undergo a scattering reaction with the carbon nuclei. {The detector rate was kept constant at a level of 60-80k events/s by sending it into a feedback loop controlling the strength of the steerers.
The beam intensity was about $10^9$ particles per cycle of 
5 minutes duration.
}
The angular distribution and energy of the scattered deuterons were recorded with the forward part of the former WASA detector (\textit{\textbf{W}ide \textbf{A}ngle \textbf{S}hower \textbf{A}pparatus}) shown in Figure~\ref{fig:WASA}, which is installed in the COSY storage ring \cite{art:wasa_at_cosy}. The WASA forward detector consists of five thick hodoscope layers composed of pizza shaped plastic scintillators for energy measurement. Each scintillator segment was read out by an individual photomultiplier tube.  A four-layer straw tube array was used to extract the angular information of each deuteron track with a resolution of $\SI{0.2}{\degree}$ in $\Theta$
and $\SI{0.5}{\degree}$ (at $\Theta=\SI{2}{\degree}$) and $\SI{2.5}{\degree}$ 
(at $\Theta=\SI{17}{\degree}$) in $\Phi$.
\cite{art:wasa_at_cosy}. An additional set of three layers of thin plastic scintillators as well as the first forward range hodoscope was used to generate the trigger signal. The signals from these four layers were discretized by discriminators for each segment and then fed into an FPGA based system that provided the trigger signal for the ADC modules. The trigger requirement was an uninterrupted track from the very first scintillator layer all the way through the first forward range hodoscope layer.
The detector covers a polar ($\Theta$) angular range from \SI{2}{\degree} to \SI{17}{\degree} and the full azimuthal ($\Phi$) range. 

This detector setup allows one to record events in bins of $\Theta$, $\Phi$ and energy of the scattered particle. The next section explains how to extract the analyzing power $A_y$ from the event rates.

\section{Analysis Method}\label{sec:analysis}
\subsection{Selection of elastic events}
\begin{figure}
    \centering
\includegraphics[width=\textwidth]{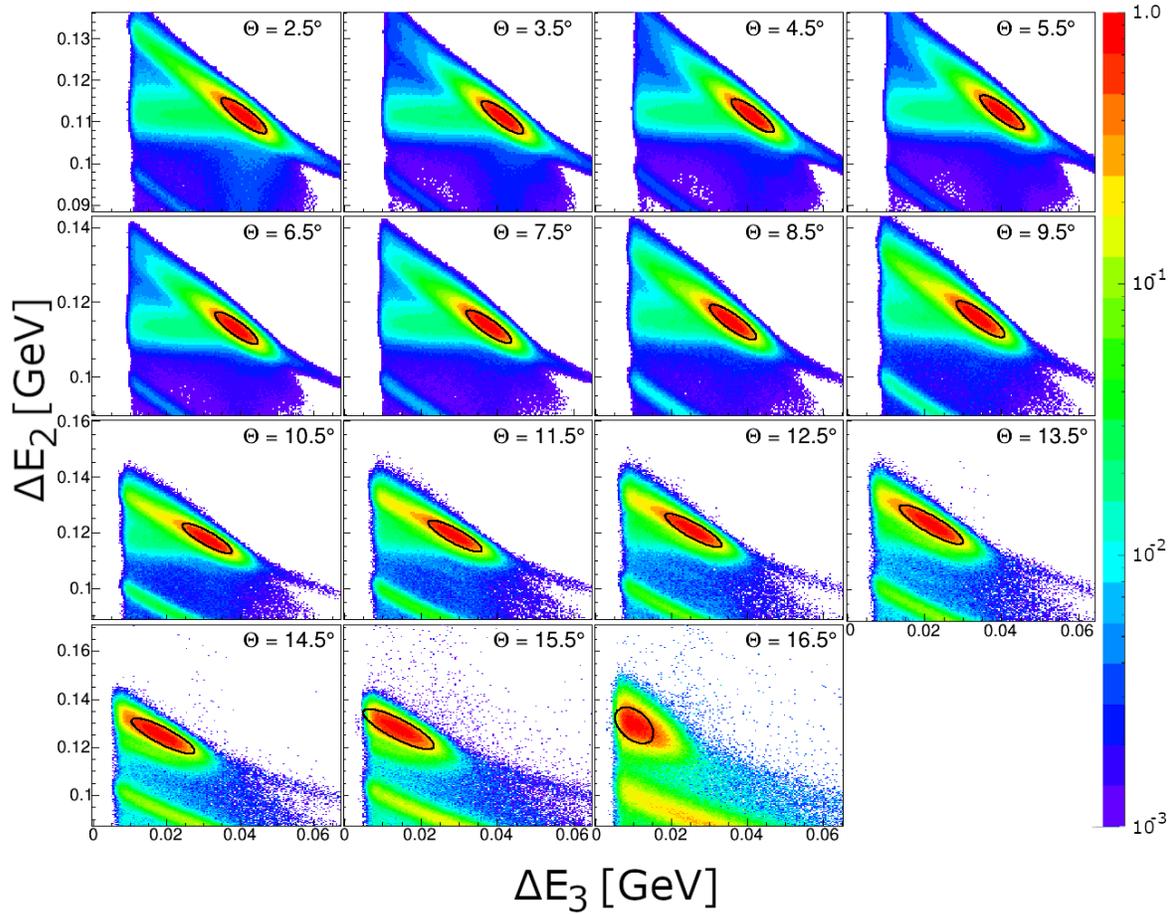}
    \caption{Energy loss in the second (FRH2) versus the third
    (FRH3) layer of 
    the forward range hodoscope, $\Delta E_2$ vs. $\Delta E_3$ 
    for various bins in $\Theta$ {for a beam energy of 270~MeV}. The centers of the one degree wide bins 
    are indicated in the plots.
    The data were fitted to a two dimensional normal distribution.
    Events within the $1\sigma$ ellipse indicated in the figure entered the analysis.
  { The diagonal areas at the bottom left of the plots originates from protons produced in a stripping reaction.
  All histograms were normalised to the their respective maximum number of events in one bin.}}
    \label{fig:eliptic_cuts}
\end{figure}
In order to select elastically scattered deuterons,
a Monte-Carlo simulation was used to perform an energy calibration of each detector layer. The multi-layer design of the detector allowed generating $\Delta E_{n-1}$ vs. $\Delta E_{n}$ plots for the stopping layer n (last reachable layer for a given energy) and the layer before (n-1) that could be used for particle identification. The elastically scattered deuterons were selected using graphical cuts on these plots (see Figure \ref{fig:eliptic_cuts}). The elastically scattered deuterons were binned in $\Delta\Theta = 1^\circ$ wide bins for the identification of the elastic events but the final analysis was performed in \SI{0.5}{\degree}-bins. 
{ The energy resolution of the forward detector is approximately  5\%. This allows to exclude events originating from break-up reactions. Note, that the contribution of the excited state 2+ (4.4 MeV) of the carbon nucleus cannot be resolved with our detector.
According to~\cite{art:satou} its contribution to the cross section at a beam energy of 270 MeV amounts to 1 -- 23\% in the angular range 2--17 degrees with an analyzing power of similar size compared to the elastic scattering.}

\subsection{Extraction of asymmetries}\label{sec:asym}
In order to extract the vector analyzing power $A_y$ from this data, it is useful to define an asymmetry parameter $\epsilon$ as follows:
\begin{equation}
    \epsilon(\Theta) = \frac{3}{2}A_y(\Theta) P_y.
    \label{eq:asymmetry}
\end{equation}

The classical way to extract $\epsilon$ would be to look at a left-right asymmetry (left corresponds to  $\Phi=0$) for two data sets with opposite vertically polarized beams~\cite{Ohlsen:1973wf}.
This requires assumptions about acceptance and integration limits  
in $\Phi$ not to dilute the sample with events where $\cos(\Phi) \approx 0$ (up and down direction). 

Here we use a method introduced in ref.~\cite{art:joerg_fabian}
which gives the best statistical accuracy and no assumptions on acceptances, except for stability in time between the data sets with different polarization states, have to be made.
The method essentially consists of weighting every event
with $\cos(\Phi)$.
This method offers multiple advantages compared to the so called cross ratio method~\cite{Ohlsen:1973wf}. It can be shown that by weighting every event $i$ by $\cos(\Phi_i)$ , the statistical error is given by:
\begin{equation}
    \sigma(\epsilon)\approx \frac{1}{\sqrt{\langle \cos^2 \Delta\Phi \rangle N}} \, ,
    \label{eq:stat_error_weighting}
\end{equation}
whereas in the cross ratio method \cite{Ohlsen:1973wf} one reaches
\begin{equation}
\sigma(\epsilon)\approx \frac{1}{\sqrt{\langle \cos \Delta\Phi \rangle^2 N}} \, .
\end{equation}
The gain in statistical error is thus
\[
  \sqrt{ \frac{\langle \cos^2 \Delta\Phi  \rangle}{\langle \cos \Delta\Phi \rangle^2}} \,  \ge 1 \, ,
\]
where $\langle \dots \rangle$ denotes the acceptance weighted average over $\Phi$.
Due to the weighting procedure the full polar range $\Delta\Phi$ from 0 to $2\pi$ can be used for the asymmetry calculation without an increase of the statistical error for larger $\Delta\Phi$. Since the events at $\Phi \approx \pm90^\circ$ are assigned with a very small weight, they cannot dilute the sample anymore. Further, this method does not require the detector acceptance $\alpha$ to be flat in $\Phi$ but it can even be determined. In addition this method reaches the Cramer-Rao bound~\cite{Barlow:213033} of minimal variance on $\epsilon$.

The analysis was performed in the following way: 
Cycles with polarization state up and down were analyzed together
with an unpolarized cycles, and asymmetries $\epsilon^\uparrow$ and $\epsilon^\downarrow$ were extracted for every $\Theta$ bin independently.
Dividing these asymmetries by the corresponding polarization 
values $P_y = P^\uparrow$ and $P_y = -P^\downarrow$ yields the analyzing power 
$A_y$.

\begin{figure}[h!]
    \centering
    \includegraphics[width=0.7\textwidth]{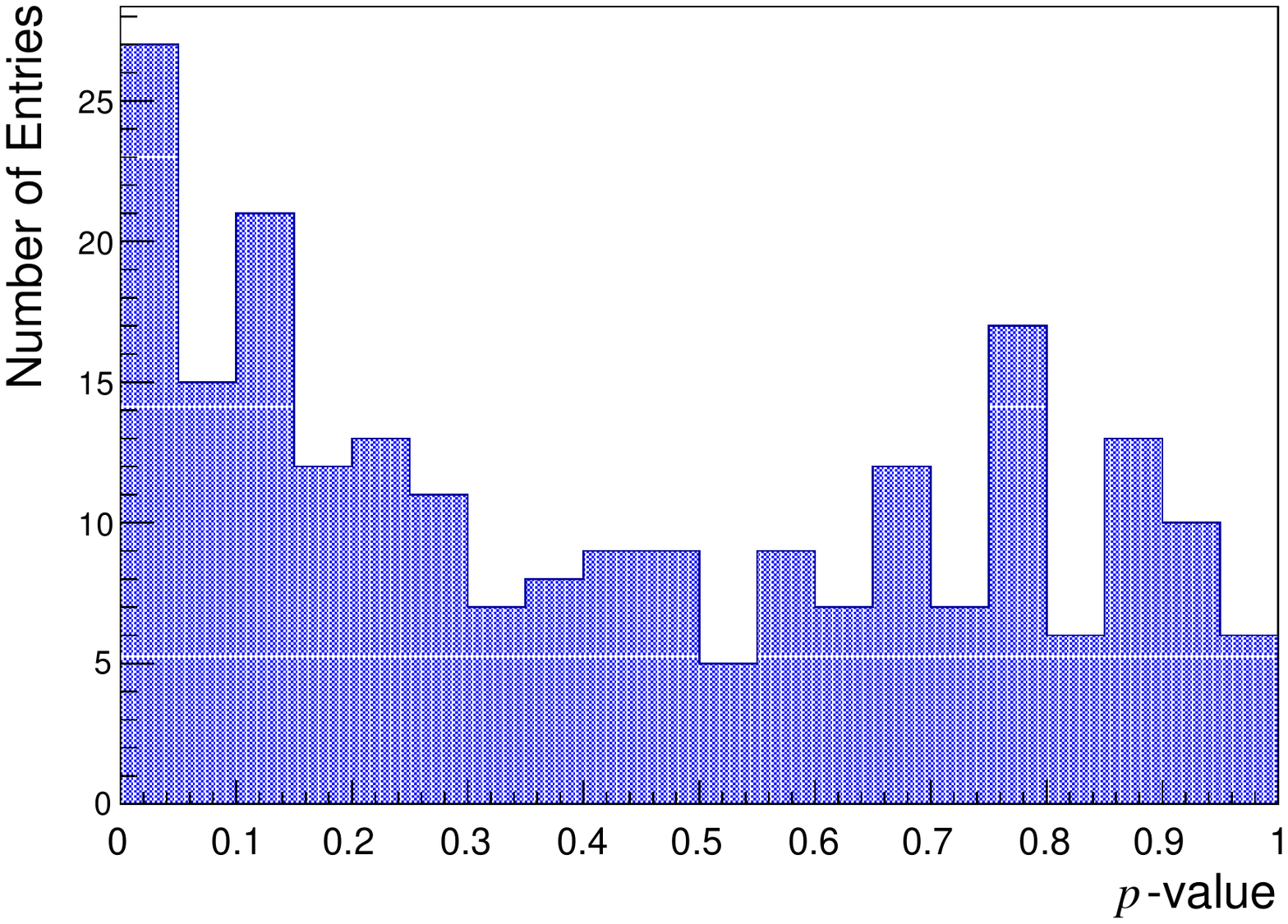}
    \caption{$p$-value distribution ($p=\int_{\chi^2}^{\infty} f(\chi^2,n) \mathrm{d}\chi^2$) for all extracted asymmetries.}
    \label{fig:chi2}
\end{figure}

The asymmetry determination involves a $\chi^2$ minimization,
(see ~\cite{art:joerg_fabian}, eq. 14), here using as input 9 measurements:
For each of the three polarization states (up, down, unpolarized) three sums
$\sum_{i=1}^{N_{\mathrm{events}}} \cos(\Phi_i)^n$ with
$n=0,1,2$ were computed. 
In total 8 parameters are extracted: three factors related to luminosity 
for the three polarization states, three factors describing the acceptance in $\Phi$ and two asymmetries $\epsilon^\uparrow$ and $\epsilon^\downarrow$.
Having 9 equations and only 8 parameters allows one to perform a $\chi^2$ test
with one degree of freedom.
Figure~\ref{fig:chi2} shows the corresponding $\chi^2$ probability
distribution which looks reasonable flat as expected. 
Since the asymmetries were calculated independently for every $\Theta$ bin
(=28), and every energy data set (in total 8),
the total number of entries is $8  \times 28 = 224$.

\subsection{Determination of $P^\uparrow$ and $P^\downarrow$}
The polarized ion source~\cite{Felden:2011nqm} used to create the polarized deuteron beam was set to produce the maximum achievable magnitude of ${2}/{3}$ pure vector polarization for deuterons. However, it was not possible to state that the desired polarization magnitude was actually reached. This was due to a failure in the low energy polarimeter which is located just after the pre-accelerator stage of COSY and normally measures the beam polarization before the main acceleration within COSY takes place.

Moreover, the asymmetry analysis of two data sets of 270~MeV deuteron measured twelve days apart, disclosed that the beam polarization was not completely stable over the whole course of the experiment. Two sets of reference deuteron carbon vector analyzing powers $A_y$ for 200~MeV \cite{art:kawabata} and for 270~MeV \cite{art:satou} are available and were used to evaluate the polarization value in our experiment. 

\begin{figure}[h!]
    \centering
    \includegraphics[width=0.75\textwidth]{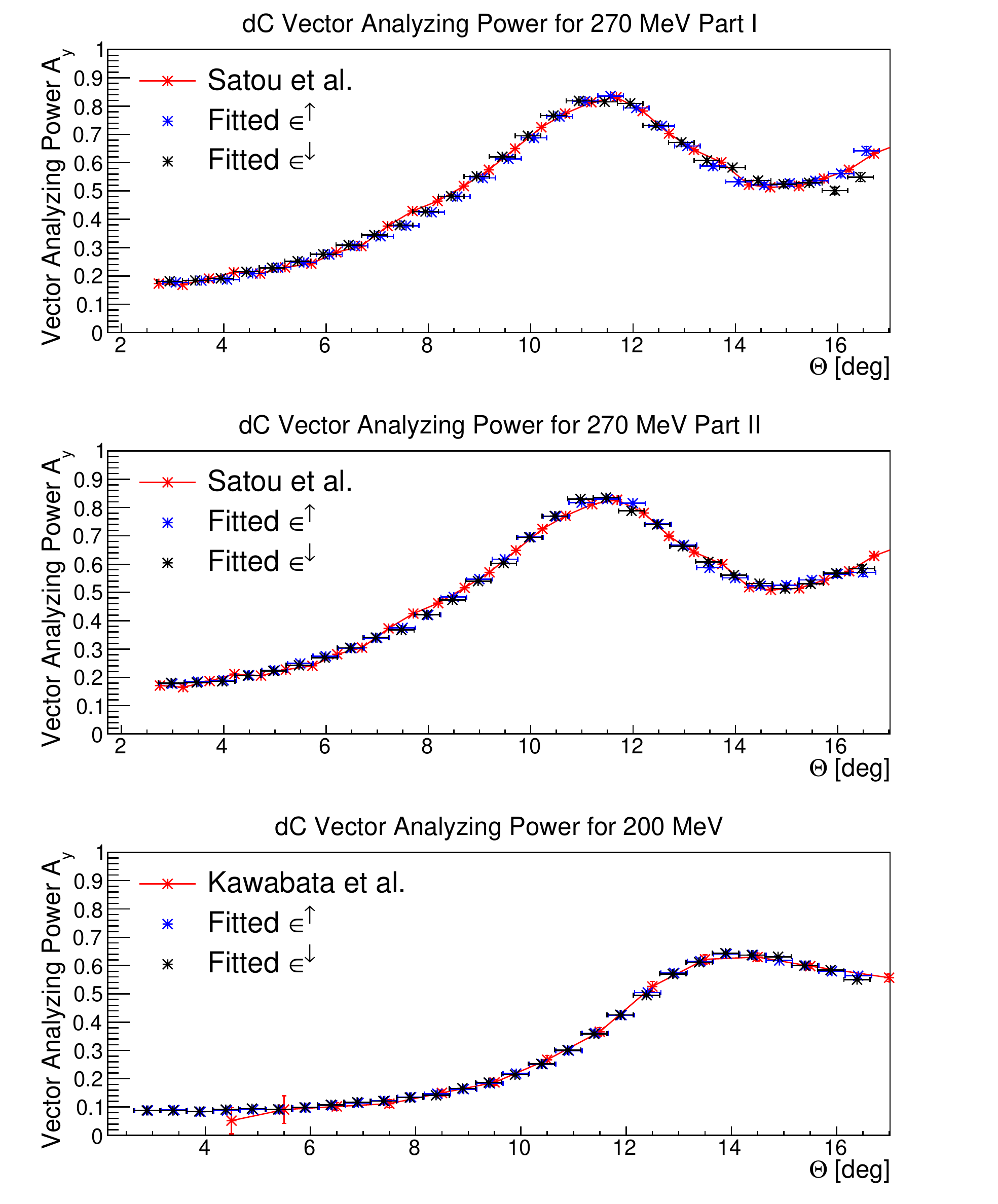}
    \caption{Measured asymmetry for the upwards  $\epsilon^\uparrow$ (blue) and downwards  $\epsilon^\downarrow$ (black) polarized beam
    as a function of the polar angle in the laboratory. The asymmetries were fitted to the reference data (red) by \textit{Satou et al.} \cite{art:satou} and \textit{Kawabata et al.} \cite{art:kawabata} to obtain the beam polarization. The 270 MeV data was measured in two sets that were fitted individually. }
    \label{fig:pol_extraction}
\end{figure}

In order to obtain polarization values $P^\uparrow$ and $P^\downarrow$,
the asymmetry parameters $\epsilon^\uparrow$ and $\epsilon^\downarrow$, calculated as described in section~\ref{sec:asym} for 200~MeV and 270~MeV, were fitted to the published vector analyzing powers at the corresponding energies. This allowed one to verify that the line-shape of our result was in agreement with the published data. From the scaling parameter of the fit, the polarization magnitude could be calculated according to Equation~(\ref{eq:asymmetry}). 
Figure~\ref{fig:pol_extraction} shows the result of these fits. Both polarization states were fitted independently, resulting in polarization value for each state. The calculated polarization magnitude is given in Table \ref{tab:pol}. The two data sets at 270~MeV were also processed separately. The line-shapes for both energies and polarization states are in a good agreement with the published data. The fit result for the comparison of our 270~MeV data with \cite{art:satou} (see Figure~\ref{fig:pol_extraction} upper two plots) resulted in polarization values which differ in the order of $\sim$\SI{10}{\percent} between the two data sets. In the result for the vector analyzing power given in \cite{art:satou}, the theoretical limit of $A_y^{max} = 1.0$ is slightly ($\sim2.5\%$) exceeded at $\Theta \approx 25\,$degree, indicating an overestimation of $A_y$. When this result is used to calculate the polarization of our beam, according to Equation~(\ref{eq:asymmetry}) this leads to an underestimation of the fitted $P_y$.
The fit result of the comparison of our 200~MeV data with \cite{art:kawabata} (see lower plot in Figure~\ref{fig:pol_extraction}) yields a polarization magnitude that is very close to the limit of pure vector polarization $P_y^{max} = {2}/{3}$. In general, it is not possible to decide which value of polarization from the three fits should be taken and, therefore, taking the average of all three results seemed to be appropriate. The average of all three data sets was found to be $P^\uparrow = $ \SI{55.4 +- 6.5}{\percent} and $P^\downarrow = $ \SI{37.4 +- 4.0}{\percent}, see Table~\ref{tab:pol}.
{ The error was taken from the unbiased estimate of the standard deviation of the three values $\sqrt{\sum_{i=1}^3 (\bar{P}-P_i)^2/(N-1)}$.}


\begin{table}[]
    \centering
  \begin{tabular}{ccc}
		\toprule
		Energy  & $P^{\uparrow}$ [\%] & $P^{\downarrow}$ [\%] \\ 
		\midrule
		200~MeV & $62.5 \pm 0.6$	  &  $41.7 \pm 0.4$   \\

		270~MeV Part I &  $53.7 \pm 0.3$ & $36.8 \pm 0.3$ \\

		270~MeV Part II &  $49.9 \pm 0.3$ & $33.7 \pm 0.2$ \\
		\midrule
		Average & $55.4 \pm 6.5$ & $37.4  \pm 4.0$ \\ 
		\bottomrule
	\end{tabular}
    \caption{Polarization obtained by comparing the asymmetries 
    $\epsilon^\uparrow$  and $\epsilon^\downarrow$ to published 
    analyzing powers $A_y$. The errors in the first three lines result from the fit in figure~\ref{fig:pol_extraction}. 
    { The last line is the the average $\bar P$ of the three measurements. The error 
    was taken from the unbiased estimate of the standard deviation of the three values $\sqrt{\sum_{i=1}^3 (\bar{P}-P_i)^2/(N-1)}$}.
    }
    \label{tab:pol}
\end{table}

\subsection{Determination of analyzing power $A_y$}

\begin{figure}[h!]
    \centering
    \includegraphics[width=0.8\textwidth,trim=0px 0px 0px 60px,clip]{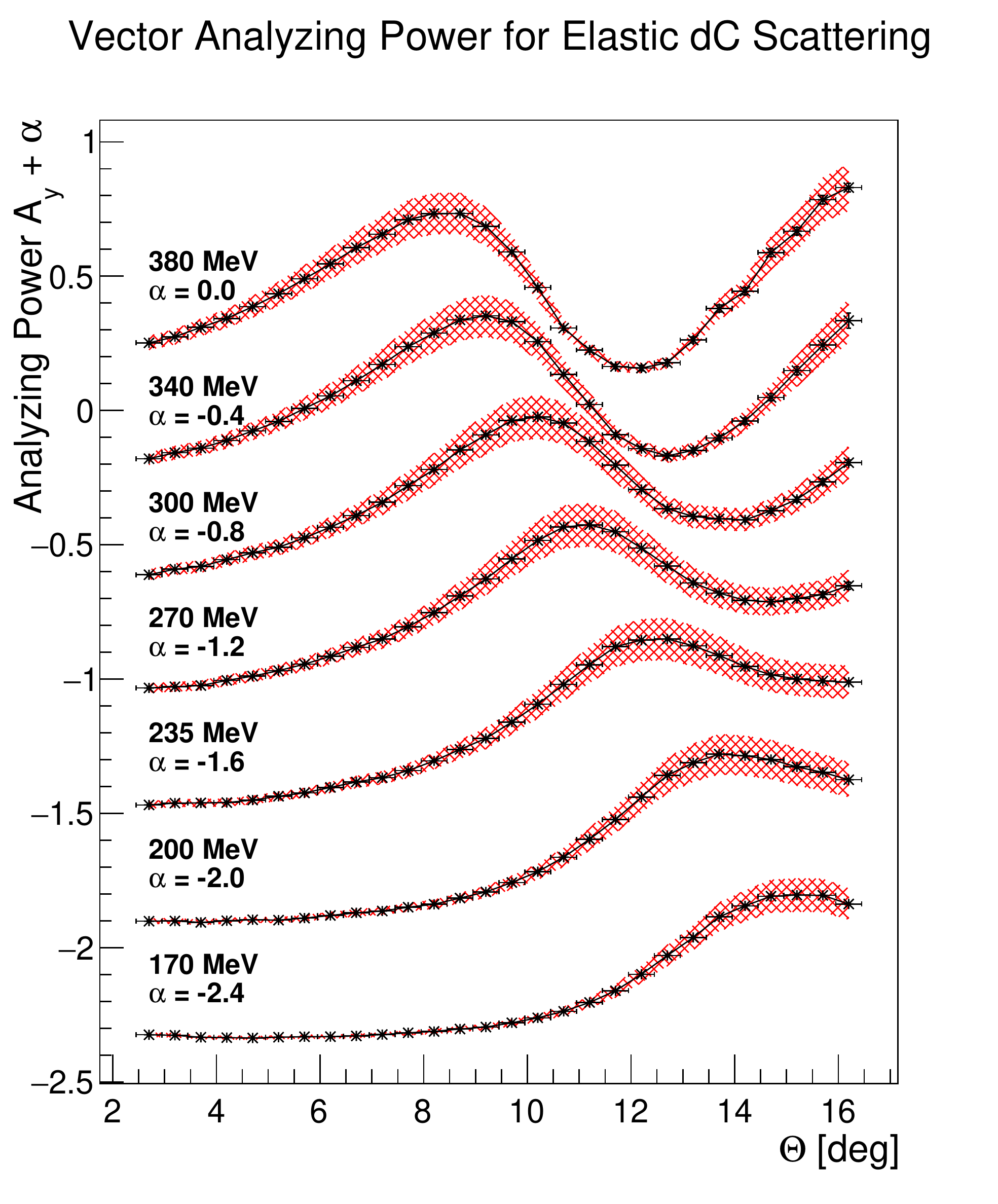}
    \caption{Reconstructed vector analyzing power for deuteron beam energies of (from top to bottom) 380~MeV, 340~MeV, 300~MeV, 270~MeV, 235~MeV, 200~MeV and 170~MeV. The curves are subsequently offset by 0.4 for better readability. The statistical errors are indicated by the vertical error bars on the data points.
    In most cases they are smaller than the symbol size. 
    The red regions show the systematic errors,
    { coming mainly from the uncertainty in the polarization measurement.}}
    \label{fig:ay}
\end{figure}

These average polarization values were now used for all beam energies
to determine the analyzing power $A_y$.
The result is shown in Figure~\ref{fig:ay} and Table~\ref{tab:ay}. The statistical error is indicated by the vertical error bars on the data points and is in most cases smaller than the symbol size.
{ The systematic error, mainly due to the error in the polarization measurement, is indicated by the red error band. 
Contributions from other sources were found to be negligible.} 
For example, fits to the data allowing 
for a tensor polarization contribution showed that $A_y$ changes by less than 1\% compared to fitting the data allowing only for vector polarization.
Other systematic uncertainties, e.g. smearing in $\Theta$ and $\Phi$, have also a negligible contribution.

\subsection{Interpretation of results}
The measurements show a maximum of $A_y$ in the  interval  of the scattering angle $\Theta=0^\circ-16^\circ$ and give some indication for the presence of  a second maximum at larger scattering angles. One should note  that two maxima are found in \cite{art:satou} at 270~MeV at $\Theta=13^\circ$ and $25^\circ$, respectively, which were fitted by the optical model \cite{art:satou}. With increasing energy the maximum observed  moves to lower scattering angles. The  maximum located at $\Theta \approx 14^\circ$ for 200~MeV is shifted to  $\Theta \approx 9^\circ$  for 380~MeV. This property is qualitatively reproduced by the three-body model of \cite{Art:IbrUz2018} formulated on the basis of  the Glauber theory with nucleon-$^{12}\mathrm{C}$ scattering amplitudes used as input. The maxima appear at the same momentum transfer $q^2$ for all seven energies.
This feature is also reproduced by the three-body model \cite{Art:IbrUz2018} and reflects the key properties of the Glauber theory which is formulated in terms of the elastic form factors of colliding nuclei.

\section{Conclusion}
The measurements performed at the COSY accelerator facility allowed for the extraction of the vector analyzing power in elastic deuteron carbon scattering reaction using a polarized deuteron beam with seven beam energies ranging from 170 to 380 MeV. These results represent a novelty with the exception of the measurements at 200~MeV \cite{art:kawabata} and 270~MeV \cite{art:satou} used as a reference. 
These results are mandatory inputs for figure of merit estimations for future polarimetry in the context of  deuteron electric dipole moment experiments planned at storage rings.  

\clearpage
\section*{Acknowledgments}
The authors would like to thank the staff of COSY for providing good working conditions and Colin Wilkin
for comments on the manuscript.  
This work has been financially supported by an ERC Advanced-Grant 
srEDM \# 694340: {\it "Electric Dipole Moments using storage rings"}
of the European Union, and by the Shota Rustaveli National Science Foundation of the Republic of Georgia (SRNSFG grant No. DI-18-298:
{\it "High precision polarimetry for charged-particle EDM searches in storage rings"}.

\begin{sidewaystable}
    \centering
    \small
  \begin{tabular}{c|ccccccc}
		\toprule
		
		
		\multirow{2}{*}{Angle [$^{\circ}$] } & \multicolumn{7}{c}{Energy [MeV]} \\
   &              170          &          200              &             235           &             270           &             300           &                340        &                380        \\
			\midrule
~2.45 - ~2.95                     & $0.076\pm 0.001$ & $0.099\pm 0.001$ & $0.131\pm 0.001$ & $0.167\pm 0.002$ & $0.188\pm 0.002$ & $0.220\pm 0.002$ & $0.252\pm 0.002$ \\
~2.95 - ~3.45                     & $0.074\pm 0.001$ & $0.100\pm 0.001$ & $0.138\pm 0.001$ & $0.171\pm 0.001$ & $0.209\pm 0.001$ & $0.243\pm 0.001$ & $0.274\pm 0.002$ \\	
~3.45 - ~3.95                     & $0.066\pm 0.001$ & $0.094\pm 0.001$ & $0.139\pm 0.001$ & $0.176\pm 0.001$ & $0.219\pm 0.001$ & $0.260\pm 0.001$ & $0.309\pm 0.002$ \\	
~3.95 - ~4.45                     & $0.065\pm 0.001$ & $0.100\pm 0.001$ & $0.140\pm 0.001$ & $0.195\pm 0.001$ & $0.245\pm 0.001$ & $0.288\pm 0.001$ & $0.343\pm 0.002$ \\	
~4.45 - ~4.95                     & $0.064\pm 0.001$ & $0.104\pm 0.001$ & $0.150\pm 0.001$ & $0.211\pm 0.001$ & $0.270\pm 0.001$ & $0.324\pm 0.001$ & $0.386\pm 0.002$ \\	
~4.95 - ~5.45                     & $0.067\pm 0.001$ & $0.103\pm 0.001$ & $0.164\pm 0.001$ & $0.231\pm 0.001$ & $0.291\pm 0.001$ & $0.359\pm 0.001$ & $0.435\pm 0.002$ \\	
~5.45 - ~5.95                     & $0.069\pm 0.001$ & $0.110\pm 0.001$ & $0.176\pm 0.001$ & $0.256\pm 0.001$ & $0.326\pm 0.001$ & $0.407\pm 0.001$ & $0.490\pm 0.002$ \\	
~5.95 - ~6.45                     & $0.069\pm 0.001$ & $0.120\pm 0.001$ & $0.197\pm 0.001$ & $0.285\pm 0.001$ & $0.366\pm 0.001$ & $0.454\pm 0.002$ & $0.546\pm 0.002$ \\	
~6.45 - ~6.95                     & $0.071\pm 0.001$ & $0.130\pm 0.001$ & $0.217\pm 0.001$ & $0.318\pm 0.001$ & $0.409\pm 0.002$ & $0.511\pm 0.002$ & $0.606\pm 0.002$ \\	
~6.95 - ~7.45                     & $0.077\pm 0.001$ & $0.137\pm 0.001$ & $0.233\pm 0.001$ & $0.350\pm 0.002$ & $0.458\pm 0.002$ & $0.571\pm 0.002$ & $0.656\pm 0.003$ \\	
~7.45 - ~7.95                     & $0.083\pm 0.001$ & $0.151\pm 0.001$ & $0.259\pm 0.001$ & $0.395\pm 0.002$ & $0.520\pm 0.002$ & $0.637\pm 0.002$ & $0.710\pm 0.003$ \\	
~7.95 - ~8.45                     & $0.088\pm 0.001$ & $0.162\pm 0.001$ & $0.296\pm 0.002$ & $0.448\pm 0.002$ & $0.580\pm 0.002$ & $0.688\pm 0.003$ & $0.733\pm 0.004$ \\	
~8.45 - ~8.95                     & $0.097\pm 0.001$ & $0.185\pm 0.001$ & $0.338\pm 0.002$ & $0.509\pm 0.002$ & $0.653\pm 0.003$ & $0.737\pm 0.003$ & $0.734\pm 0.004$ \\	
~8.95 - ~9.45                     & $0.105\pm 0.001$ & $0.208\pm 0.002$ & $0.379\pm 0.002$ & $0.573\pm 0.003$ & $0.709\pm 0.003$ & $0.752\pm 0.004$ & $0.686\pm 0.005$ \\	
~9.45 - ~9.95                     & $0.120\pm 0.002$ & $0.243\pm 0.002$ & $0.440\pm 0.002$ & $0.647\pm 0.003$ & $0.762\pm 0.004$ & $0.731\pm 0.005$ & $0.590\pm 0.006$ \\	
~9.95 - 10.45                     & $0.139\pm 0.002$ & $0.283\pm 0.002$ & $0.506\pm 0.003$ & $0.716\pm 0.004$ & $0.776\pm 0.004$ & $0.656\pm 0.005$ & $0.458\pm 0.007$ \\
10.45 - 10.95                     & $0.163\pm 0.002$ & $0.337\pm 0.002$ & $0.580\pm 0.003$ & $0.766\pm 0.004$ & $0.753\pm 0.005$ & $0.535\pm 0.006$ & $0.307\pm 0.008$ \\
10.95 - 11.45                     & $0.196\pm 0.002$ & $0.404\pm 0.003$ & $0.653\pm 0.003$ & $0.774\pm 0.005$ & $0.684\pm 0.006$ & $0.422\pm 0.007$ & $0.225\pm 0.009$ \\
11.45 - 11.95                     & $0.240\pm 0.003$ & $0.477\pm 0.003$ & $0.721\pm 0.004$ & $0.749\pm 0.005$ & $0.597\pm 0.006$ & $0.309\pm 0.008$ & $0.164\pm 0.010$ \\
11.95 - 12.45                     & $0.301\pm 0.003$ & $0.561\pm 0.003$ & $0.745\pm 0.004$ & $0.688\pm 0.006$ & $0.506\pm 0.007$ & $0.258\pm 0.008$ & $0.158\pm 0.010$ \\
12.45 - 12.95                     & $0.371\pm 0.003$ & $0.642\pm 0.004$ & $0.749\pm 0.005$ & $0.621\pm 0.006$ & $0.434\pm 0.007$ & $0.231\pm 0.009$ & $0.177\pm 0.012$ \\  
12.95 - 13.45                     & $0.438\pm 0.003$ & $0.688\pm 0.004$ & $0.724\pm 0.005$ & $0.558\pm 0.006$ & $0.405\pm 0.008$ & $0.252\pm 0.009$ & $0.262\pm 0.012$ \\
13.45 - 13.95                     & $0.515\pm 0.004$ & $0.720\pm 0.004$ & $0.688\pm 0.005$ & $0.520\pm 0.007$ & $0.397\pm 0.008$ & $0.298\pm 0.011$ & $0.380\pm 0.013$ \\
13.95 - 14.45                     & $0.556\pm 0.004$ & $0.715\pm 0.004$ & $0.647\pm 0.005$ & $0.493\pm 0.007$ & $0.393\pm 0.009$ & $0.361\pm 0.011$ & $0.444\pm 0.013$ \\
14.45 - 14.95                     & $0.592\pm 0.004$ & $0.700\pm 0.004$ & $0.616\pm 0.006$ & $0.488\pm 0.008$ & $0.427\pm 0.009$ & $0.449\pm 0.013$ & $0.588\pm 0.014$ \\
14.95 - 15.45                     & $0.596\pm 0.004$ & $0.673\pm 0.004$ & $0.600\pm 0.006$ & $0.500\pm 0.008$ & $0.469\pm 0.010$ & $0.548\pm 0.015$ & $0.667\pm 0.015$ \\
15.45 - 15.95                     & $0.595\pm 0.004$ & $0.653\pm 0.005$ & $0.593\pm 0.006$ & $0.514\pm 0.010$ & $0.534\pm 0.011$ & $0.643\pm 0.019$ & $0.784\pm 0.016$ \\
15.95 - 16.45                     & $0.563\pm 0.004$ & $0.625\pm 0.005$ & $0.588\pm 0.006$ & $0.548\pm 0.013$ & $0.606\pm 0.010$ & $0.735\pm 0.028$ & $0.830\pm 0.016$ \\
		\bottomrule
	\end{tabular}
    \caption{Vector analyzing power for all seven beam energies in the format 
    \mbox{$A_y \pm \sigma_{stat.}$}. The systematic uncertainty amounts to \SI{10}{\percent}.}
    \label{tab:ay}
\end{sidewaystable}

\newpage
\bibliographystyle{unsrt}
\bibliography{references}{}

\end{document}

%% file: authors.tex

\author[1,2]{F.~M\"uller}

\affil[1]{Institut f\"ur Kernphysik, Forschungszentrum J\"ulich, 52425 J\"ulich, Germany}
\affil[2]{III. Physikalisches Institut B, RWTH Aachen University, 52056 Aachen, Germany}

\author[1]{M.~\.{Z}urek}

%

%


%

%
%
\author[1]{Z.~Bagdasarian}

\author[3]{L.~Barion}
\affil[3]{University of Ferrara and INFN, 44100 Ferrara, Italy}
\author[4]{M.~Berz}
\affil[4]{Department of Physics and Astronomy, Michigan State University,  East Lansing, Michigan 48824, USA}
%
%
\author[5]{I.~Ciepal}
\affil[5]{Institute of Nuclear, Physics Polish Academy of Sciences, 31342 Crakow, Poland}
\author[3]{G.~Ciullo}

\author[1,6]{S.~Dymov}
\affil[6]{Laboratory of Nuclear Problems, Joint Institute for Nuclear Research, 141980 Dubna, Russia}
\author[2]{D. Eversmann}

\author[2]{M.~Gaisser}

\author[1]{R.~Gebel}

\author[1,2]{K.~Grigoryev}

\author[1]{D.~Grzonka}

%
%
\author[1]{V.~Hejny}

\author[1,2]{N.~Hempelmann}

\author[1]{J.~Hetzel}
%
\author[1,2]{F.~Hinder}
%
\author[1]{A.~Kacharava}

\author[1]{V.~Kamerdzhiev}
%
\author[1]{I.~Keshelashvili}

\author[7]{I.~Koop}
\affil[7]{Budker Institute of Nuclear Physics, 630090 Novosibirsk, Russia}
\author[6]{A.~Kulikov}
%
\author[1,8]{A.~Lehrach}
\affil[8]{JARA--FAME (Forces and Matter Experiments), Forschungszentrum J\"ulich and RWTH Aachen University, Germany}
\author[3]{P.~Lenisa}

\author[9]{N.~Lomidze}
\affil[9]{High Energy Physics Institute, Tbilisi State University, 0186 Tbilisi, Georgia}

%
\author[1]{B.~Lorentz}
%
\author[2]{P.~Maanen}
%
\author[9]{G.~Macharashvili}
%
\author[10]{A.~Magiera}
\affil[10]{Institute of Physics, Jagiellonian University, 30348 Cracow, Poland}
\author[9]{D.~Mchedlishvili}

\author[1]{A.~Nass}
%
\author[11,12]{N.N. Nikolaev}
\affil[11]{L.D. Landau Institute for Theoretical Physics, 142432 Chernogolovka, Russia}
\affil[12]{Moscow Institute for Physics and Technology, 141700 Dolgoprudny, Russia}
%
%
\author[3]{A.~Pesce}
%
\author[1]{D.~Prasuhn}
%
\author[1,2,8,*]{J.~Pretz}

\author[1]{F.~Rathmann}
%
\author[3]{V. Rolando}

\author[1,2]{M.~Rosenthal}
%

\author[1]{A.~Saleev}
%
\author[1,2]{V.~Schmidt}
%
%
\author[1]{Y.~Senichev}
%
\author[9]{D. Shergelashvili}

\author[1,6]{V.~Shmakova}
%
\author[13,14]{A.~Silenko}
\affil[13]{Research Institute for Nuclear Problems, Belarusian State University, 220030 Minsk, Belarus}
\affil[14]{Bogoliubov Laboratory of Theoretical Physics, Joint Institute for Nuclear Research, 141980 Dubna, Russia}
\author[15]{J.~Slim}
\affil[15]{Institut f\"ur Hochfrequenztechnik, RWTH Aachen University, 52056 Aachen, Germany}
\author[16]{H.~Soltner}
\affil[16]{Zentralinstitut f\"ur Engineering, Elektronik und Analytik (ZEA-1), Forschungszentrum J\"ulich, 52425 J\"ulich, Germany}
\author[2]{A.~Stahl}
%
\author[1]{R.~Stassen}
%
\author[17]{E.~Stephenson}
\affil[17]{Indiana University Center for Spacetime Symmetries, Bloomington,  Indiana 47405, USA}
%
%
\author[1,8]{H.~Str\"oher}
%
\author[9]{M.~Tabidze}
%
\author[18]{G.~Tagliente}
\affil[18]{INFN, 70125 Bari, Italy}
\author[19]{R.~Talman}
\affil[19]{Cornell University, Ithaca,  New York 14850, USA}
%
%
\author[1,2]{F.~Trinkel}
%
\author[6,20,21]{Yu.~Uzikov}
\affil[20]{Dubna State University, 141980 Dubna, Russia}
\affil[21]{Department of Physics, M.V. Lomonosov Moscow State University, 119991 Moscow, Russia}
\author[1]{Yu.~Valdau}
%
\author[4]{E.~Valetov}
%
%
\author[1]{C.~Weidemann}
%
\author[10]{A.~Wro\'{n}ska}
%
\author[22]{P.~W\"ustner}
\affil[22]{Zentralinstitut f\"ur Engineering, Elektronik und Analytik (ZEA-2), Forschungszentrum J\"ulich, 52425 J\"ulich, Germany}

\affil[*]{corresponding author \texttt{pretz@physik.rwth-aachen.de}}